\documentclass[a4paper,11pt]{article}
\usepackage{jcappub}

\usepackage{slashed}
\usepackage{youngtab}

\newcommand\fverb{\setbox\fverbbox=\hbox\bgroup\verb}
\newcommand\fverbdo{\egroup\medskip\noindent%
            \fbox{\unhbox\fverbbox}\ }
\newcommand\fverbit{\egroup\item[\fbox{\unhbox\fverbbox}]}
\newbox\fverbbox
\newcommand{\nablaslash}{\not{\hbox{\kern-3pt $\nabla$}}}

\title{Inflaton decay and  reheating in  nonminimal derivative coupling}

\author{Yun~Soo~Myung}
\author{and Taeyoon~Moon}
\affiliation{Institute of Basic Science and Department of Computer
Simulation, Inje University,\\
Gimhae 621-749, Korea}

\emailAdd{ysmyung@inje.ac.kr} \emailAdd{tymoon@inje.ac.kr}

\abstract{We investigate the inflaton decay and reheating period
after the end of inflation in the non-minimal derivative coupling
(NDC) model with chaotic potential. In general, this model is known
to provide an enhanced slow-roll inflation caused by gravitationally
enhanced friction. We find  violent oscillations of Hubble parameter
which induces oscillations of the sound speed squared, implying the
Lagrangian instability of curvature perturbation $\zeta$ under the
comoving gauge $\varphi=0$. Also, it is shown  that the curvature
perturbation   blows up at $\dot{\phi}=0$, leading to the breakdown
of the comoving gauge at $\dot{\phi}=0$. Therefore, we use the
Newtonian gauge to perform the perturbation analysis where the
Newtonian potential is employed as a physical variable. The
curvature perturbation is not considered as a physical variable
which describes a relevant perturbation during reheating. }

\begin{document}

\maketitle \flushbottom

\section{Introduction}
It is known that reheating is a crucial  epoch which connects
inflation to the hot big-bang phase~\cite{Turner:1983he}. This era
is conceptually very important, but it is observationally poorly
known. The physics of this phase transition is thought to be highly
non-linear~\cite{Amin:2014eta}.  Also, the physics of reheating has
turned out to be very
complicated~\cite{Traschen:1990sw,Kofman:1994rk,Kofman:1996mv,Kofman:1997yn}.
Since the first CMB constraints have performed on the reheating
temperature by the WMAP7~\cite{Martin:2010kz},  the current Planck
satellite measurements of the CMB anisotropy constrain the kinematic
properties of the reheating era for  almost 200  of the inflationary
models~\cite{Martin:2014nya}.

 The nonminimal derivative coupling
(NDC)~\cite{Amendola:1993uh,Sushkov:2009hk}  was made  by coupling
the inflaton kinetic term to the Einstein tensor such that the
friction is enhanced gravitationally~\cite{Germani:2010gm}. The
gravitationally enhanced friction mechanism has been  considered as
an alternative  to increase friction of an inflaton rolling down its
own potential.
 Actually, the  NDC  makes a steep (non-flat) potential adequate for inflation without
introducing    higher-time derivative terms (ghost
state)~\cite{Germani:2011ua,Germani:2011mx}. This implies that  the
NDC increases friction and thus, it flattens
 the potential effectively.

It is worth to note that  there was a difference in whole dynamics
between
 canonical coupling (CC) and NDC  even for taking the  same potential~\cite{Myung:2015tga}.
A clear difference appears after the end of inflation. We note  that
there are three phases in the CC case~\cite{Donoghue:2007ze}: i)
Initially, kinetic energy dominates. ii) Due to the rapid decrease
of the kinetic energy, the trajectory runs into the inflationary
attractor line (potential energy dominated). All initial
trajectories are attracted to this line, which is the key feature of
slow-roll inflation. iii) At the end of inflation, the inflaton
velocity decreases.  Then, there is inflaton decay and reheating
[the appearance of spiral sink in the phase portrait
$(\phi,\dot{\phi})$].

 On the other hand, three stages of  NDC are
as follows: i) Initially, potential energy dominates. ii) Due to the
gravitationally enhanced friction (restriction on inflaton velocity
$\dot{\phi}$), all initial trajectories are attracted quickly to the
inflationary attractor. iii) At the end of inflation, the inflaton
velocity increases. Then, there is inflaton decay  and followed by
reheating. Importantly, there exist oscillations of inflaton
velocity without damping due to violent oscillations of Hubble
parameter. This provides stable limited cycles in the phase portrait
$(\phi,\dot{\phi})$, instead of spiral sink in CC.   However, it was
shown that analytic expressions for inflaton and Hubble parameter
after the inflation could be found by applying the averaging method
to the NDC~\cite{Ghalee:2013ada}. The inflaton oscillates with
time-dependent frequency, while the Hubble parameter does not
oscillate. Introducing an interacting Lagrangian of ${\cal L}_{\rm
int}=-\frac{1}{2}g^2\phi^2\chi^2$, they have claimed that the
parametric resonance instability is absent, implying  a crucial
difference when comparing to the CC. This requires a complete
solution by solving NDC-equations numerically. Recently, the authors
in~\cite{Ema:2015oaa} have investigated particle production after
inflation by considering  the combined model of CC+NDC. They have
insisted that the violent oscillation of Hubble parameter causes
particle production even though the Lagrangian instability appears
due to oscillations of the sound speed squared $c_s^2$ which also
appeared  in the generalized Galilean theory~\cite{Ohashi:2012wf}.

One usually assumes  that the field mode is frozen
(time-independent) at late time after entering into the
super-horizon. Therefore, it was accepted that the perturbation
during the reheating is less important than that of inflation.
However, in exploring  the effects of reheating on the cosmological
perturbations of CC case, one has to face the breakdown of the
curvature perturbation $\zeta$ at $\dot{\phi}=0$ when choosing the
comoving gauge of $\varphi=0$. This issue  may be bypassed by
replacing $\dot{\phi}^2$ by its time average $\langle
\dot{\phi}^2\rangle$ over the inflaton
oscillation~\cite{Finelli:1998bu,Jedamzik:2010dq,Easther:2010mr}.
Recently, it was proposed that the breakdown of the comoving gauge
$\varphi=0$ at $\dot{\phi}=0$ could be resolved by introducing the
$cd$-gauge which eliminates $\varphi$ in the Hamiltonian formalism
of the CC model and thus, provides a well-behaved curvature
perturbation $\zeta$~\cite{Algan:2015xca}. However, it turned out
that choosing the Newtonian gauge is necessary to study the
perturbation during the oscillating period, since the comoving gauge
is not suitable for performing the perturbation analysis during the
reheating~\cite{Germani:2015plv}.

In this work, we find a complete solution for inflaton and Hubble
parameter by solving the NDC-equations numerically in Section 2. The
NDC model may be dangerous because the inflaton becomes strongly
coupled when the Hubble parameter tends towards zero. Hence, we wish
to obtain a complete solution for inflaton and Hubble parameter by
solving the CC+NDC-equations numerically in Section 3. Here,  we can
control mutual importance of the CC and NDC  by adjusting two
coefficients. In Section 4, we  will investigate the curvature
perturbation $\zeta$ during reheating by considering  the NDC  with
the chaotic potential and choosing the comoving gauge. We find that
violent oscillations of Hubble parameter
 induce oscillations of the sound speed squared, implying the
Lagrangian instability of curvature perturbation.  More seriously,
we show that the curvature perturbation blows up at $\dot{\phi}=0$,
implying that the curvature perturbation is ill-defined under the
comoving gauge of $\varphi=0$.  This suggests a different gauge
without problems at  $\dot{\phi}=0$. Hence, we choose the Newtonian
gauge to perform the perturbation analysis where the Newtonian
potential is considered  as a physical variable in Section 5.

%%%%%%%%%%%%%%%%%%%%%%%%%%%%%%%%%%%%%%%%%%%%%%%%%%%%%%%%%%%%%%%%%%%%%
%%%%%%%%%%%%%%%%%%%%%%%%%%%%%%%%%%%%%%%%%%%%%%%%%%%%%%%%%%%%%%%%%%%%%
%%%%%%%%%%%%%%%%%%%%%%%%%%%%%%%%%%%%%%%%%%%%%%%%%%%%%%%%%%%%%%%%%%%%%
\section{NDC with chaotic potential}
%%%%%%%%%%%%%%%%%%%%%%%%%%%%%%%%%%%%%%%%%%%%%%%%%%%%%%%%%%%%%%%%%%%%%
%%%%%%%%%%%%%%%%%%%%%%%%%%%%%%%%%%%%%%%%%%%%%%%%%%%%%%%%%%%%%%%%%%%%%
%%%%%%%%%%%%%%%%%%%%%%%%%%%%%%%%%%%%%%%%%%%%%%%%%%%%%%%%%%%%%%%%%%%%%

We introduce  an inflation model  including the NDC of scalar field
$\phi$ with the chaotic potential~\cite{Feng:2014tka,Myung:2015tga}
\begin{eqnarray} \label{mact}
S_{\rm}=\frac{1}{2}\int d^4x \sqrt{-g}\Big[M_{\rm
P}^2R+\frac{1}{\tilde{M}^2}G_{\mu\nu}\partial^{\mu}\phi\partial^{\nu}\phi-2V(\phi)\Big],~~V=V_0\phi^2,
\end{eqnarray}
where $M_{\rm P}$ is a reduced Planck mass, $\tilde{M}$ is a mass
parameter and  $G_{\mu\nu}$ is the Einstein tensor. Here, we do not
include  a canonical coupling (CC) term like as a conventional
combination of CC+NDC
[$(g_{\mu\nu}-G_{\mu\nu}/\tilde{M}^2)\partial^\mu\phi\partial^\nu
\phi$]~\cite{Tsujikawa:2012mk,Skugoreva:2013ooa} because this
combination won't   make the whole analysis transparent.

From the action (\ref{mact}), we derive  the Einstein and inflaton
equations
\begin{eqnarray} \label{einseq}
&&G_{\mu\nu} =\frac{1}{M_{\rm P}^2} T_{\mu\nu},\\
\label{einseq-1}&&\frac{1}{\tilde{M}^2}G^{\mu\nu}\nabla_{\mu}\nabla_{\nu}\phi+V'=0,
\end{eqnarray}
where $T_{\mu\nu}$ takes a complicated  form
\begin{eqnarray}
 T_{\mu\nu}&=&\frac{1}{\tilde{M}^2}\Big[\frac{1}{2}R\nabla_{\mu}\phi\nabla_{\nu}\phi
-2\nabla_{\rho}\phi\nabla_{(\mu}\phi R_{\nu)}^{\rho}
+\frac{1}{2}G_{\mu\nu}(\nabla\phi)^2-R_{\mu\rho\nu\sigma}\nabla^{\rho}\phi\nabla^{\sigma}\phi
\nonumber
\\&&\hspace*{5em}-\nabla_{\mu}\nabla^{\rho}\phi\nabla_{\nu}\nabla_{\rho}\phi
+(\nabla_{\mu}\nabla_{\nu}\phi)\nabla^2\phi\nonumber\\
&&
\label{em1}-g_{\mu\nu}\Big(-R^{\rho\sigma}\nabla_{\rho}\phi\nabla_{\sigma}\phi+\frac{1}{2}(\nabla^2\phi)^2
-\frac{1}{2}(\nabla^{\rho}\nabla^{\sigma}\phi)\nabla_{\rho}\nabla_{\sigma}\phi
\Big)\Big].
\end{eqnarray}
Considering a  flat FRW spacetime  by introducing cosmic time $t$ as
\begin{eqnarray} \label{deds1}
ds^2_{\rm FRW}~=~\bar{g}_{\mu\nu}dx^\mu
dx^\nu~=~-dt^2+a^2(t)\delta_{ij}dx^idx^j, \label{deds2}
\end{eqnarray}
two Friedmann
 and inflaton   equations (NDC-equations) derived from (\ref{einseq}) and (\ref{einseq-1}) are given
by
\begin{eqnarray}
H^2&=&\frac{1}{3M_{\rm P}^2}\Big[\frac{9H^2}{2\tilde{M}^2}\dot{\phi}^2+V\Big],\label{Heq}\\
&&\nonumber\\
\dot{H}&=&-\frac{1}{2M_{\rm
P}^2}\Big[\dot{\phi}^2\Big(\frac{3H^2}{\tilde{M}^2}-\frac{\dot{H}}{\tilde{M}^2}\Big)-\frac{2H}{\tilde{M}^2}\dot{\phi}\ddot{\phi}
\Big],\label{dHeq}\\
&&\nonumber\\
&&\hspace*{-4em}\frac{3H^2}{\tilde{M}^2}\ddot{\phi}+3H\Big(\frac{3H^2}{\tilde{M}^2}+\frac{2\dot{H}}{\tilde{M}^2}\Big)\dot{\phi}+V'=0.\label{seq}
\end{eqnarray}
Here $H=\dot{a}/a$ is the Hubble parameter and  the overdot
($\dot{}$) denotes derivative with respect to time $t$. It is
evident from (\ref{Heq}) that the energy density for the NDC is
positive (ghost-free).
%%%%%%%%%%%%%%%%%%%%%%%%%%%%%%%%%%%%%%%%%%%%%%%%%%%%%%%%%%%%%%%%%
%%%%%%%%%%%%%%%%%%%%%%%%%%%%%%%%%%%%%%%%%%%%%%%%%%%%%%%%%%%%%%%%%
%%%%%%%%%%%%%%%%%%%%%%%%%%%%%%%%%%%%%%%%%%%%%%%%%%%%%%%%%%%%%%%%%
\begin{figure}[t!]
\begin{center}
\begin{tabular}{cc}
\includegraphics[width=.9
\linewidth,origin=tl]{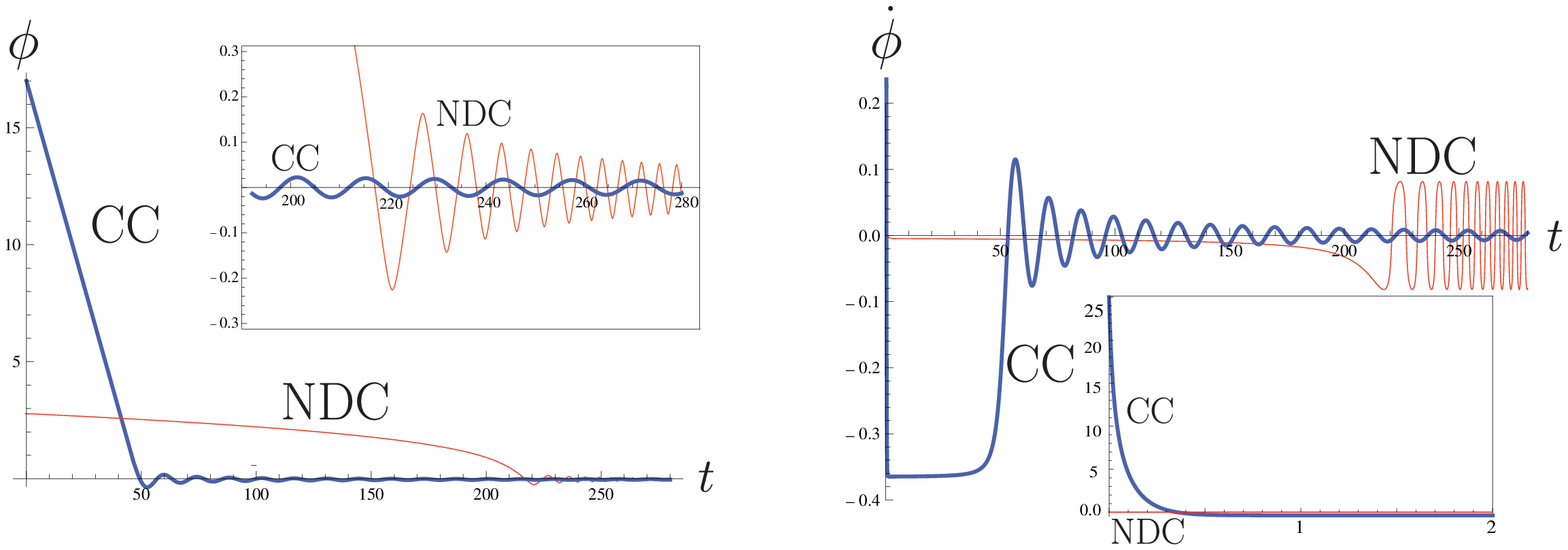}
\end{tabular}
\end{center}
\caption{The whole evolution of $\phi(t)$ [left] and $\dot{\phi}(t)$
[right] with respect to time $t$ for chaotic potential $V=V_0\phi^2$
with $V_0=0.1$. The left figure shows that the inflaton varies
little during large inflationary period ($0\le t\le 200$) for the
NDC, while it varies quickly during small inflationary period ($0\le
t\le 45$) for the CC. After inflation (see figure in box), $\phi$
decays with oscillation for CC, while it oscillates rapidly for NDC.
The right one indicates that for large $t$, $\dot{\phi}$ oscillates
without damping for  NDC, while it oscillates with damping for the
CC. Figure in box shows  initially kinetic energy  phase for CC and
initially potential phase for NDC.  }
\end{figure}

At this stage, the CC  model of $-g_{\mu\nu}\partial^\mu \phi
\partial^\nu \phi$ is introduced  to compare with the NDC case. In
this case, the CC-equations are given by
\begin{eqnarray}
H^2&=&\frac{1}{3M_{\rm P}^2}\Big[\frac{1}{2}\dot{\phi}^2+V\Big],\label{Heqc}\\
\dot{H}&=&-\frac{1}{2M_{\rm
P}^2}\dot{\phi}^2,\label{dHeqc}\\
\ddot{\phi}&+&3H\dot{\phi}+m^2\phi=0 \label{seqc}
\end{eqnarray}
with $m^2=2V_0$.  Fig. 1  shows a whole evolution of $\phi$ and
$\dot{\phi}$  based on numerical computation. When the universe
evolves according to (\ref{Heqc})-(\ref{seqc}), there are three
phases in the CC case~\cite{Donoghue:2007ze}: i) Initially, kinetic
energy dominates [see Fig. 1 (right)]. ii) Due to the rapid decrease
of the kinetic energy, the trajectory runs quickly to the
inflationary attractor line. All initial trajectories are attracted
to this line, which is the key feature of slow-roll inflation. iii)
Finally, after the end of inflation, there is inflaton decay and
reheating which corresponds to spiral sink in the phase portrait
($\phi,\dot{\phi}$). Explicitly, (\ref{Heqc}) can be parameterized
by using the Hubble parameter $H$ and the angular variable $\theta$
as
\begin{eqnarray} \label{hat1}
\dot{\phi}&=&\sqrt{6}HM_{\rm P}\cos \theta,\\
\label{hat2}m\phi&=&\sqrt{6}HM_{\rm P}\sin \theta,
\end{eqnarray}
while (\ref{dHeqc}) and (\ref{seqc}) implies
\begin{eqnarray}\label{hat3}
\dot{H}&=&-3H^2\cos^2
\theta,\\
\label{hat4}\dot{\theta}&=&-m-\frac{3}{2}H\sin(2\theta).
\end{eqnarray}
For $m\gg H$, (\ref{hat4}) reduces to  $\dot{\theta} \simeq -m$
 which implies a solution of $\theta\simeq-mt$. Plugging the latter  into
(\ref{hat2}) indicates that $\phi$ oscillates with  frequency
$\omega\simeq m=0.45$ for $V_0=0.1$. Solving (\ref{hat3}) leads to
\begin{equation} \label{hat5}
H(t)\simeq\frac{2}{3t}\Big[1+\frac{\sin(2mt)}{2mt}\Big]^{-1}
\end{equation}
which shows small oscillations around $\frac{2}{3t}$. Actually, its
time rate is given by
\begin{equation} \label{hat6}
\dot{H}(t)\simeq-\frac{16m^2 \cos^2(mt)}{3\Big[2mt
+\sin(2mt)\Big]^2}=-\frac{8m^2[1+ \cos(2mt)]}{3\Big[2mt
+\sin(2mt)\Big]^2}
\end{equation}
whose amplitude approaches zero ($-\frac{2}{3t^2}$) with
oscillations as $t$ increases. Its frequency is given by
 $\omega^{\rm CC}_{\dot{H}}=2m$. Substituting (\ref{hat5}) into (\ref{hat2}) provides us the scalar
\begin{equation}\label{hat7}
\phi(t)\simeq\sqrt{\frac{8}{3}}\frac{M_{\rm
P}}{mt}\sin(mt)\Bigg[1-\frac{\sin(2mt)}{2mt}\Bigg],
\end{equation}
which implies that after the end of inflation,  the friction becomes
subdominant and thus,  $\phi(t)$ becomes an oscillator whose
amplitude gets damped due to the universe evolution $H$. The  time
rate is given by
\begin{equation}\label{hat8}
\dot{\phi}(t)\simeq\sqrt{\frac{8}{3}}\frac{M_{\rm
P}}{t}\cos(mt)\Bigg[1-\frac{\sin(2mt)}{2mt}\Bigg].
\end{equation}
We observe that $\omega^{\rm CC}_\phi=\omega^{\rm
CC}_{\dot{\phi}}=m$.

 The scale factor can be extracted from (\ref{hat5})
as
\begin{equation}
a(t)\simeq t^{\frac{2}{3}},
\end{equation}
while the energy density of $\phi$ decreases in the same way as the
energy density of non-relativistic particles of mass $m$
\begin{equation}
\rho_{\phi}=\frac{1}{2}\Big[\dot{\phi}^2+m^2\phi^2\Big] \sim
\frac{1}{a^3}. \end{equation} This indicates that the inflaton
oscillations can be interpreted to be a collection of scalar
particles, which are independent from each other, oscillating
coherently at the same frequency $m$.

Differing with the CC model, the upper limit of $\dot{\phi}^2$ is
set for the NDC model
\begin{eqnarray} \label{cond-NDC}
0 <\dot{\phi}^2 \le \phi_{c}^2\equiv\frac{2}{3}M_{\rm
P}^2\tilde{M}^2,
\end{eqnarray}
which comes from Eq.(\ref{Heq}) showing that $3M^2_{\rm
P}H^2(1-\dot{\phi}^2/\phi^2_c)=V\ge0$.  Based on
(\ref{Heq})-(\ref{seq}), we can figure out a whole picture
numerically [see Fig.1 (left)]. Three stages are in the NDC: i)
Initially, potential energy dominates. ii) Due to the
gravitationally enhanced friction, all initial trajectories  are
attracted quickly to the inflationary attractor. iii) At the end of
inflation, the inflaton velocity increases. Then, there is inflaton
decay  and followed by reheating. However,  there exist oscillations
of inflaton velocity without damping. This provides stable limited
cycles in the phase portrait $(\phi,\dot{\phi})$, instead of spiral
sink. We stress that   an analytic solution for NDC is  not yet
known  because equations (\ref{Heq})-(\ref{seq}) are too complicated
to be solved. However, the would-be analytic solution might be found
in~\cite{Ghalee:2013ada}.

%%%%%%%%%%%%%%%%%%%%%%%%%%%%%%%%%%%%%%%%%%%%%%%%%%%%%%%%%%%%%%%%%
%%%%%%%%%%%%%%%%%%%%%%%%%%%%%%%%%%%%%%%%%%%%%%%%%%%%%%%%%%%%%%%%%
%%%%%%%%%%%%%%%%%%%%%%%%%%%%%%%%%%%%%%%%%%%%%%%%%%%%%%%%%%%%%%%%%
\begin{figure}[t!]
\begin{center}
\begin{tabular}{c}
\includegraphics[width=.90\linewidth,origin=tl]{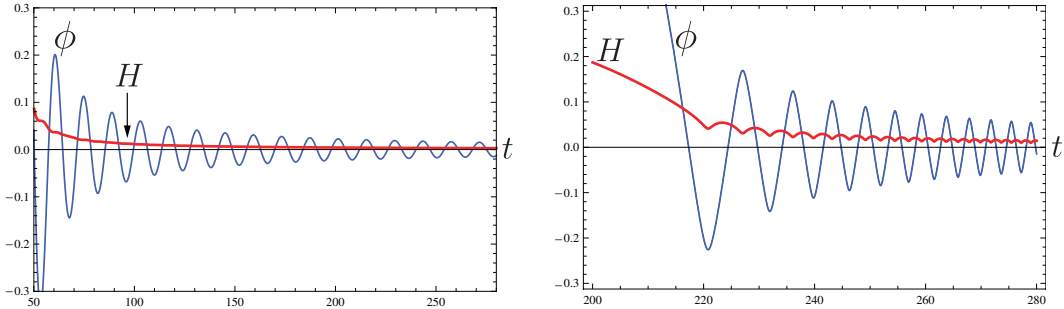}
\end{tabular}
\end{center}
\caption{After the end of inflation, behaviors of inflation $\phi$
(blue) and Hubble parameter $H$ (red)  with respect to time $t$.
Left picture is for CC while right one represents the NDC case. We
observe violent oscillations of $H$ for NDC. Here, angular frequency
of $H$ is given by $\omega_H(t)=2\omega_\phi(t)$ for NDC, while
frequency of $\phi$ is $\omega^{\rm CC}_{\phi}=m$ for CC. }
\end{figure}
%%%%%%%%%%%%%%%%%%%%%%%%%%%%%%%%%%%%%%%%%%%%%%%%%%%%%%%%%%%%%%%%%
Now we are in a position to focus on the reheating period after the
end of inflation (post-inflationary phase). We remind the reader
that the friction term dominates in the slow-roll inflation period,
while the friction term becomes subdominant in the reheating
process. Therefore,  the inflaton becomes an oscillator  whose
amplitude gets damped due to the universe expansion. Fig. 2 shows
behaviors of inflation $\phi$ and Hubble parameter $H$  with respect
to time $t$. Left figure is designed for CC [(\ref{hat7}) and
(\ref{hat5})], while the right one represents the NDC case. We
observe violent oscillations of $H$ for NDC. Here, oscillation
frequency  of $H$  is given by $\omega_H(t)=2\omega_\phi(t)$ for
NDC. Fig. 3 indicates behaviors of inflaton velocity $\dot{\phi}$
(blue) and Hubble parameter $H$ (red)  with respect to time $t$.
Left picture is for CC [(\ref{hat8}) and (\ref{hat5})], while the
right one represents the NDC case. We find violent oscillations of
$H$ for NDC. Here, oscillation frequency  of $H$  is still given by
$\omega_H(t)=2\omega_{\dot{\phi}}(t)$ for  NDC. Importantly, we
observe a sizable difference that $\dot{\phi}$ oscillates with
damping (CC), while it oscillates without damping and its frequency
increases (NDC).
%%%%%%%%%%%%%%%%%%%%%%%%%%%%%%%%%%%%%%%%%%%%%%%%%%%%%%%%%%%%%%%%%
%%%%%%%%%%%%%%%%%%%%%%%%%%%%%%%%%%%%%%%%%%%%%%%%%%%%%%%%%%%%%%%%%

\begin{figure}[t!]
\begin{center}
\begin{tabular}{c}
\includegraphics[width=.90\linewidth,origin=tl]{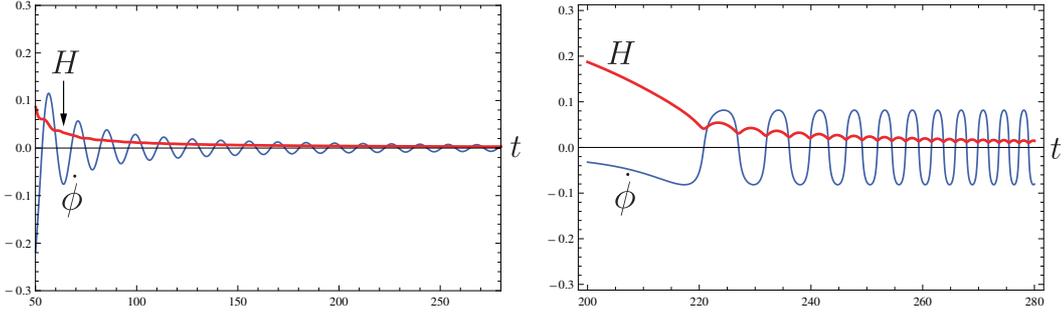}
\end{tabular}
\end{center}
\caption{After the end of inflation, behaviors of inflaton velocity
$\dot{\phi}$ (blue) and Hubble parameter $H$ (red)  with respect to
time $t$. Left picture is for CC, while right one represents the NDC
case. We observe violent oscillations of $H$ for NDC. Oscillation
frequency of $H$  is given by $\omega_H(t)=2\omega_{\dot{\phi}}(t)$
for NDC, while the frequency of $\dot{\phi}$ is $\omega^{\rm
CC}_{\dot{\phi}}=m$ for CC.}
\end{figure}
%%%%%%%%%%%%%%%%%%%%%%%%%%%%%%%%%%%%%%%%%%%%%%%%%%%%%%%%%%%%%%%%
%%%%%%%%%%%%%%%%%%%%%%%%%%%%%%%%%%%%%%%%%%%%%%%%%%%%%%%%%%%%%%%%%
%%%%%%%%%%%%%%%%%%%%%%%%%%%%%%%%%%%%%%%%%%%%%%%%%%%%%%%%%%%%%%%%%
%%%%%%%%%%%%%%%%%%%%%%%%%%%%%%%%%%%%%%%%%%%%%%%%%%%%%%%%%%%%%%%%%
\begin{figure}[t!]
\begin{center}
\begin{tabular}{c}
\includegraphics[width=.90\linewidth,origin=tl]{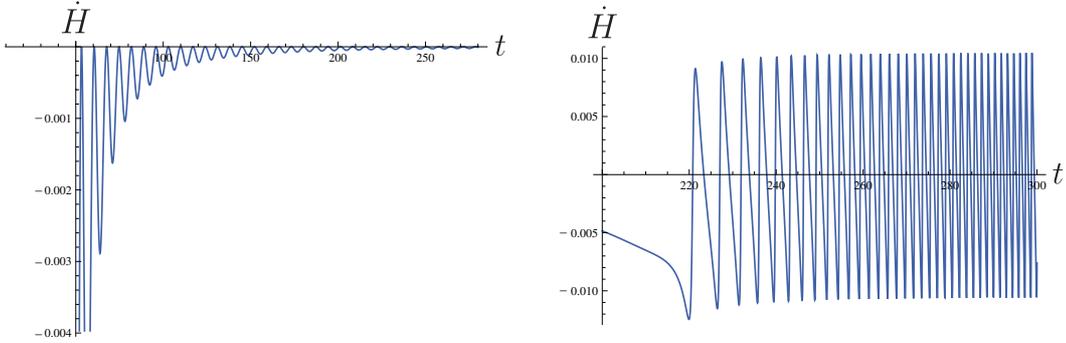}
\end{tabular}
\end{center}
\caption{Oscillations of $\dot{H}$ after the end of inflation: Left
(CC) and Right (NDC). Here we observe the difference between CC and
NDC: $\dot{H}\le 0$ for CC
 and $-0.01 \le \dot{H} \le 0.01$.}
\end{figure}
%%%%%%%%%%%%%%%%%%%%%%%%%%%%%%%%%%%%%%%%%%%%%%%%%%%%%%%%%%%%%%%%%
Here, we  mention that different behaviors of $\phi$ and
$\dot{\phi}$ between CC and NDC have arisen from  different
oscillations of their Hubble parameter $H$. Their change of rates
$\dot{H}$ are depicted in Fig. 4, which would be used to obtain the
sound speed squared $c^2_s$. It is quite interesting to note   the
difference that $\dot{H}$ of CC [(\ref{hat6})] approaches zero
(along $-\frac{2}{3t^2}$) with oscillations ($\omega^{\rm
CC}_{\dot{H}}=2m$), while $\dot{H}$ oscillates between $-0.01$ and
0.01 with frequency $\omega_{\dot{H}}(t)$.

 At this stage, we note  that
an analytic solution might  be obtained by using  the averaging
method~\cite{Ghalee:2013ada}. This is given by
\begin{eqnarray} \label{analtic-sol2}
&&H_{\rm a}(t)=\frac{2}{3(2-\sqrt{2})t},\\
\label{analtic-sol3}&&\phi_{\rm a}(t)=\frac{\sqrt{6}M_{\rm P}H_{\rm
a}(t)}{ m}
\cos\Big[\frac{m\tilde{M}}{2}(2-\sqrt{2})(\sqrt{2}-\frac{1}{2})t^2\Big],
\end{eqnarray}
where $\phi_{\rm a}(t)$ oscillates with time-dependent frequency.
Their time-rates are given by
\begin{eqnarray}
&&\dot{H}_{\rm a}(t)=-\frac{2}{3(2-\sqrt{2})t^2},\\
&&\dot{\phi}_{\rm a}(t)=-\frac{(4-\sqrt{2})M_{\rm
P}\tilde{M}}{\sqrt{3}}\sin\Big[\frac{m\tilde{M}}{2}(2-\sqrt{2})(\sqrt{2}-\frac{1}{2})t^2\Big]+\cdots.
\end{eqnarray}
We wish to comment here that even though $\dot{\phi}_{\rm a}(t)$
could mimic $\dot{\phi}$ in the right picture of Fig. 3,
$\dot{H}_{\rm a}(t)$ could not describe oscillations of $\dot{H}$ in
the right-picture of Fig. 4. This implies that the analytic solution
(\ref{analtic-sol2}) is not a proper solution to  NDC-equations
because $H_{\rm a}(t)$ did not show violent oscillations of Hubble
parameter. Also we observe  the difference in frequency  between CC
and NDC: $\omega^{\rm CC}_{\phi}=\omega^{\rm CC}_{\dot{\phi}}=m$, $
\omega^{\rm CC}_{\dot{H}}=2m$ (time-independent) and
$\omega_H(t)=2\omega_{\phi}=2\omega_{\dot{\phi}}$,
$\omega_{\dot{H}}(t)$(time-dependent).

Hence, it is not proven  that the parametric resonance is absent for
NDC when considering the decay of inflaton into a relativistic field
(${\cal L}_{\rm int}=-\frac{1}{2}g^2\phi^2\chi^2$), whereas the
parametric resonance is present for CC.

%%%%%%%%%%%%%%%%%%%%%%%%%%%%%%%%%%%%%%%%%%%%%%%%%%%%%%%%%%%%%%%%%
%%%%%%%%%%%%%%%%%%%%%%%%%%%%%%%%%%%%%%%%%%%%%%%%%%%%%%%%%%%%%%%%%
\section{CC + NDC with chaotic potential}

%%%%%%%%%%%%%%%%%%%%%%%%%%%%%%%%%%%%%%%%%%%%%%%%%%%%%%%%%%%%%%%%%
%%%%%%%%%%%%%%%%%%%%%%%%%%%%%%%%%%%%%%%%%%%%%%%%%%%%%%%%%%%%%%%%%
\begin{figure}[t!]
\begin{center}
\begin{tabular}{c}
\includegraphics[width=.90\linewidth,origin=tl]{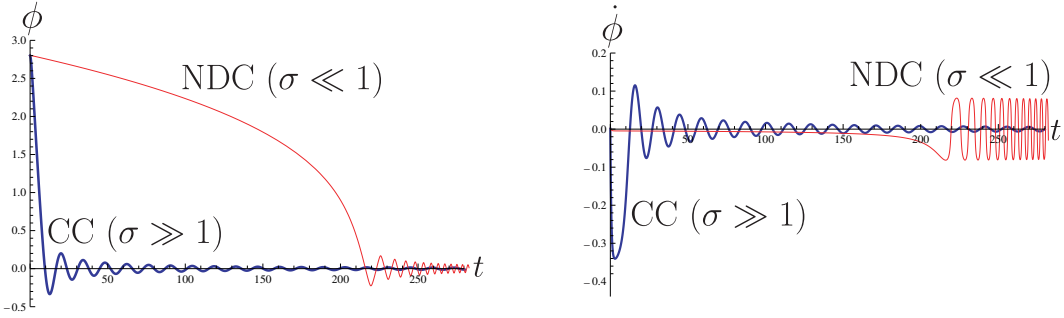}
\end{tabular}
\end{center}
\caption{The whole evolution of $\phi(t)$ [left] and $\dot{\phi}(t)$
[right] with respect to time $t$ for chaotic potential $V=V_0\phi^2$
with $V_0=0.1$. In these figures, CC-dominant (blue) and
NDC-dominant (red) cases correspond to $\sigma=10^4\gg1$ and
$\sigma=10^{-4}\ll1$, respectively. }
\end{figure}
%%%%%%%%%%%%%%%%%%%%%%%%%%%%%%%%%%%%%%%%%%%%%%%%%%%%%%%%%%%%%%%%%
%%%%%%%%%%%%%%%%%%%%%%%%%%%%%%%%%%%%%%%%%%%%%%%%%%%%%%%%%%%%%%%%%
%%%%%%%%%%%%%%%%%%%%%%%%%%%%%%%%%%%%%%%%%%%%%%%%%%%%%%%%%%%%%%%%%
\begin{figure}[t!]
\begin{center}
\begin{tabular}{c}
\includegraphics[width=.90\linewidth,origin=tl]{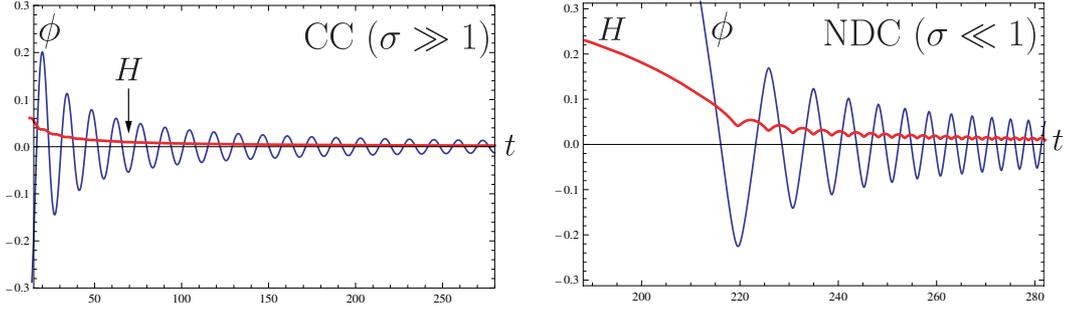}
\end{tabular}
\end{center}
\caption{After the end of inflation, behaviors of inflaton $\phi$
(blue) and Hubble parameter $H$ (red)  with respect to time $t$.
Left picture is for CC-dominant case ($\sigma=10^{4}\gg1$), while
right one represents the NDC-dominant ($\sigma=10^{-4}\ll1$) case.}
\end{figure}
%%%%%%%%%%%%%%%%%%%%%%%%%%%%%%%%%%%%%%%%%%%%%%%%%%%%%%%%%%%%%%%%%
%%%%%%%%%%%%%%%%%%%%%%%%%%%%%%%%%%%%%%%%%%%%%%%%%%%%%%%%%%%%%%%%%
%%%%%%%%%%%%%%%%%%%%%%%%%%%%%%%%%%%%%%%%%%%%%%%%%%%%%%%%%%%%%%%%%
%%%%%%%%%%%%%%%%%%%%%%%%%%%%%%%%%%%%%%%%%%%%%%%%%%%%%%%%%%%%%%%%%
\begin{figure}[t!]
\begin{center}
\begin{tabular}{c}
\includegraphics[width=.90\linewidth,origin=tl]{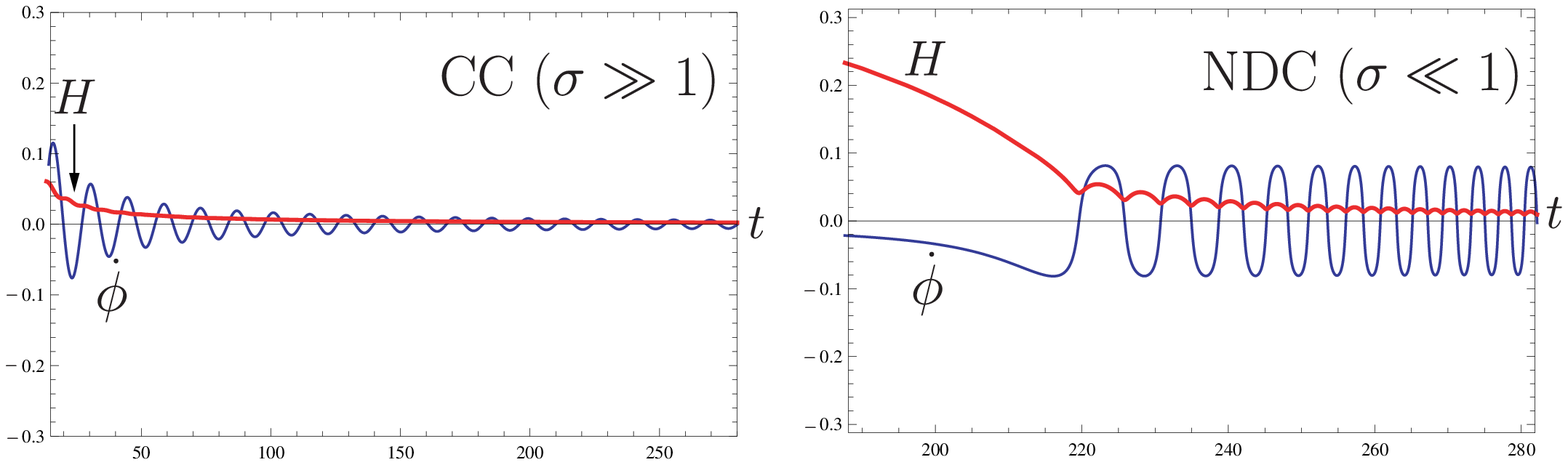}
\end{tabular}
\end{center}
\caption{After the end of inflation, behaviors of inflaton velocity
$\dot{\phi}$ (blue) and Hubble parameter $H$ (red). Left picture is
for CC-diminant case ($\sigma=10^{4}\gg1$), while right one
represents the NDC-dominant ($\sigma=10^{-4}\ll1$) case.}
\end{figure}
In this section, we wish to study the homogeneous evolution of the
CC$+$NDC model. It is noted  that the  NDC (\ref{mact}) without CC
term might be dangerous when the Hubble parameter tends to zero.
That is,  tending of Hubble parameter to zero  may induce a strongly
coupled inflaton\footnote{We thank the anonymous referee for
pointing out this.}. To this end, we start with the CC$+$NDC action
with  chaotic potential as
\begin{eqnarray} \label{mactt}
S_{\rm CC+NDC}=\frac{1}{2}\int d^4x \sqrt{-g}\Big[M_{\rm
P}^2R-\Big(\sigma_{\rm C}g_{\mu\nu}-\sigma_{\rm
N}G_{\mu\nu}\Big)\partial^{\mu}\phi\partial^{\nu}\phi-2V(\phi)\Big],~V=V_0\phi^2.
\end{eqnarray}
The CC$+$NDC-equations are given by
\begin{eqnarray}
H^2&=&\frac{1}{3M_{\rm P}^2}\Big[\frac{1}{2}(\sigma_{\rm C}+9H^2\sigma_{\rm N})\dot{\phi}^2+V\Big],\label{Heqf}\\
&&\nonumber\\
\dot{H}&=&-\frac{1}{2M_{\rm P}^2}\Big[\dot{\phi}^2(\sigma_{\rm
C}+3H^2\sigma_{\rm N}-\dot{H}\sigma_{\rm N})-2H\sigma_{\rm
N}\dot{\phi}\ddot{\phi}
\Big],\label{dHeqf}\\
&&\nonumber\\
&&\hspace*{-4em}(\sigma_{\rm C}+3H^2\sigma_{\rm
N})\ddot{\phi}+3H(\sigma_{\rm C}+3H^2\sigma_{\rm
N}+2\dot{H}\sigma_{\rm N})\dot{\phi}+V'=0\label{seqf},
\end{eqnarray}
where $\sigma_{\rm N}=1/\tilde{M}^2$ and $\sigma_{\rm C}$ is
introduced to denote a new coefficient for the CC term.

Now we can solve Eqs.(\ref{Heqf})-(\ref{seqf}) numerically. Denoting
$\sigma\equiv \sigma_{\rm C}/\sigma_{\rm N}$, we obtain  the
CC-dominant case  by taking $\sigma\gg1$ and the NDC-dominant case
by taking $\sigma\ll1$.  Fig. 5 shows  the whole evolution of $\phi$
(left) and $\dot\phi$ (right), while Fig. 6 and 7 indicate  the
evolution after the end of inflation for ($\phi,H)$ and
($\dot\phi,H)$, respectively. Also, after the end of inflation,
$\dot{H}(t)$ is depicted in Fig. 8.

Importantly, we note that the evolutions given in Fig. 1-4
correspond to those in Fig. 5-8, respectively. We observe  that they
are very similar to each other. Therefore, it is clear  that the
evolution of the NDC-equations (\ref{Heq})-(\ref{seq}) could be
recovered from the NDC-dominant case of the  CC$+$NDC-equations
(\ref{Heqf})-(\ref{seqf}), while the CC-equations
(\ref{Heqc})-(\ref{seqc}) could be recovered from the CC-dominant
case of the CC$+$NDC-equations.
%%%%%%%%%%%%%%%%%%%%%%%%%%%%%%%%%%%%%%%%%%%%%%%%%%%%%%%%%%%%%%%%%
%%%%%%%%%%%%%%%%%%%%%%%%%%%%%%%%%%%%%%%%%%%%%%%%%%%%%%%%%%%%%%%%%
\begin{figure}[t!]
\begin{center}
\begin{tabular}{c}
\includegraphics[width=.90\linewidth,origin=tl]{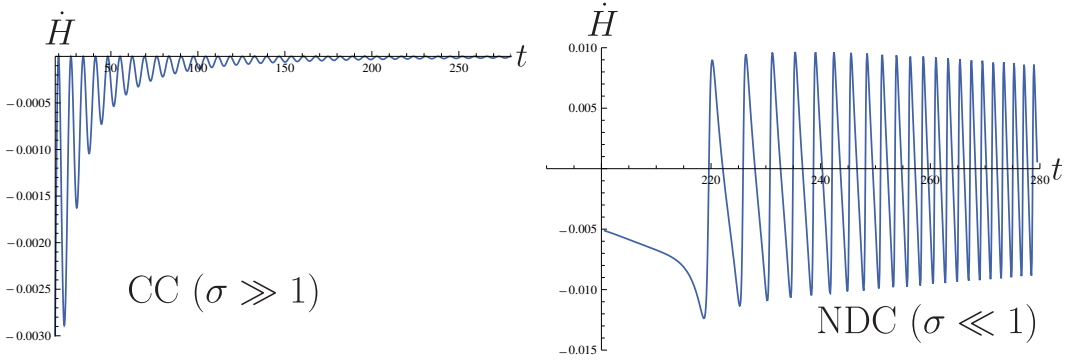}
\end{tabular}
\end{center}
\caption{Oscillations of $\dot{H}$ after the end of inflation: Left
(CC-dominance:~$\sigma=10^{4}\gg1$) and right
(NDC-dominance:~$\sigma=10^{-4}\ll1$). }
\end{figure}
%%%%%%%%%%%%%%%%%%%%%%%%%%%%%%%%%%%%%%%%%%%%%%%%%%%%%%%%%%%%%%%%%
\section{Curvature perturbation in the comoving gauge }
In order to find what happens in the post-inflationary phase, it
would be better  to analyze the perturbation.  We use the ADM
formalism to resolve the mixing between scalar of metric and
inflaton
\begin{eqnarray}
ds_{\rm ADM}^2=-N^2dt^2+\gamma_{ij}(dx^i+\beta^i dt)(dx^j+\beta^j
dt),
\end{eqnarray}
where $N$, $\beta_i$, and $\gamma_{ij}$ denote  lapse, shift vector,
and spatial metric tensor. In this case, the action (\ref{mact}) can
be written as
\begin{eqnarray}\label{mact1}
S=\int d^4x\sqrt{-g}\Big[\frac{M_{\rm
P}^2}{2}R+\frac{G^{00}}{\tilde{M}^2}\frac{\dot{\phi}^2}{2}-V\Big],
\end{eqnarray}
where
\begin{eqnarray}
R&=&R^{(3)}+\frac{1}{N^2}(E^{ij}E_{ij}-E^2)-2\nabla_{\mu}(Kn^{\mu})-\frac{2}{N}\Delta^{(3)}N,\\
G^{00}&=&\frac{1}{2N^2}\Big[R^{(3)}+\frac{1}{N^2}(E^2-E^{ij}E_{ij})\Big].
\end{eqnarray}
Here $E_{ij}$ is related to the extrinsic curvature $K_{ij}$  and
$n^\mu$ is the unit normal vector of the timelike hypersurface as
\begin{eqnarray}
E_{ij}=NK_{ij}=\frac{1}{2}(\nabla_{i}^{(3)}\beta_j+\nabla_{j}^{(3)}\beta_i-\dot{\gamma}_{ij}),~~n^{\mu}=\frac{1}{N}(1,-\beta^i).
\end{eqnarray}
Then, we express the action (\ref{mact1}) as
\begin{eqnarray}\label{mact2}
S&=&\frac{M_{\rm P}^2}{2}\int d^4x\sqrt{\gamma}\Big[R^{(3)}\Big(N+\frac{\dot{\phi}^2}{2NM_{\rm P}^2\tilde{M}^2}\Big)\nonumber\\
&&\hspace*{5em}+(E^{ij}E_{ij}-E^2)\Big(\frac{1}{N}-\frac{\dot{\phi}^2}{2N^3M_{\rm
P}^2\tilde{M}^2}\Big)-\frac{2NV}{M_{\rm P}^2}\Big].
\end{eqnarray}
Varying  (\ref{mact2}) with respect to $N$ and $\beta_j$ lead to two
constraints
\begin{eqnarray}
&&\hspace*{-3em}R^{(3)}\Big(1-\frac{\dot{\phi}^2}{2N^2M_{\rm P}^2\tilde{M}^2}\Big)-(E^{ij}E_{ij}-E^2)\Big(\frac{1}{N^2}-\frac{3\dot{\phi}^2}{2N^4M_{\rm P}^2\tilde{M}^2}\Big)-\frac{2V}{M_{\rm P}^2}=0,\label{hceq}\\
&&\hspace*{10em}\nabla_i^{(3)}\Big[\Big(\frac{1}{N}-\frac{\dot{\phi}^2}{2N^3M_{\rm
P}^2\tilde{M}^2}\Big)(E^{i}_j-\delta^{i}_j E)\Big]=0.\label{mceq}
\end{eqnarray}
Hereafter, we choose the comoving  gauge for the inflaton
($\phi=\phi(t)+\varphi$)
\begin{equation}
\label{comoving-g}\varphi=0.
\end{equation}
 For simplicity, we consider the scalar perturbations
\begin{equation}
N=1+\alpha,~\beta_i=\partial_i\psi,~\gamma_{ij}=a^2e^{2\zeta}\delta_{ij},
\end{equation}
where $\zeta$ denotes the curvature perturbation. Solving
(\ref{hceq}) and (\ref{mceq}), we find the perturbed relations
\begin{eqnarray}
\alpha=\frac{A_1}{H}\dot{\zeta},~~\psi=-\frac{A_1}{H}\zeta+\chi,~~\partial_i^2\chi=\frac{a^2}{H^2}\frac{A_1^2
A_2}{1-\epsilon_{N}^H/3}\dot{\zeta},
\end{eqnarray}
where $A_{1,2}$ and a slow-roll parameter $\epsilon_N^H$ are given
by
\begin{eqnarray}\label{a12}
A_1=\frac{1-\epsilon_N^H/3}{1-\epsilon_N^H},~~A_2=\frac{\epsilon_N^H
H^2(1+\epsilon_N^H)}{1-\epsilon_N^H/3},~~\epsilon_N^H=\frac{3\dot{\phi}^2}{2M_{\rm
P}^2\tilde{M}^2}.
\end{eqnarray}
Now we wish to expand (\ref{mact2}) to second order to obtain its
bilinear action. Making some integration by parts, we find the
bilinear  action for $\zeta$ as
\begin{eqnarray}\label{s2f}
\delta S_{(2)}=M_{\rm P}^2\int d^4x a^3\frac{A_1^2
A_2}{H^2}\Big[\dot{\zeta}^2-\frac{c_s^2}
{a^2}(\partial_i\zeta)^2\Big].
\end{eqnarray}
Here the sound speed squared $c_s^2$ is given by
\begin{eqnarray}
c_s^2&=&\frac{H^2}{A_1^2A_2}A_3\label{csss}\\
\label{sss}&=&1+\frac{4}{9A_1}\frac{\epsilon_N^H}{1+\epsilon_N^H}+\frac{2\dot{H}}{H^2}\frac{1-\epsilon_N^H/3}{1+\epsilon_N^H}
\end{eqnarray}
with
\begin{eqnarray}\label{a3e}
A_3=\frac{d}{adt}\Big[\frac{aA_1}{H}\Big(1-\frac{\epsilon_N^H}{3}\Big)\Big]-1-\frac{\epsilon_N^H}{3}.
\end{eqnarray}
We have $A_1^2 A_2/H^2\ge0$, which means ghost-free. Unfortunately,
we find from Fig. 9 that $c_s^2$ (NDC) oscillates increasingly after
the end of inflation, while it is constant for CC. The former has
arisen from the presence of $\dot{H}$ in (\ref{sss}) and it may
induce the Lagrangian instability (gradient instability) which leads
to the fact that the curvature perturbation $\zeta$ grows
violently~\cite{Ema:2015oaa}.
%%%%%%%%%%%%%%%%%%%%%%%%%%%%%%%%%%%%%%%%%%%%%%%%%%%%%%%%%%%%%%%%%
%%%%%%%%%%%%%%%%%%%%%%%%%%%%%%%%%%%%%%%%%%%%%%%%%%%%%%%%%%%%%%%%%
%%%%%%%%%%%%%%%%%%%%%%%%%%%%%%%%%%%%%%%%%%%%%%%%%%%%%%%%%%%%%%%%%
\begin{figure}[t!]
\begin{center}
\begin{tabular}{cc}
\includegraphics[width=.9
\linewidth,origin=tl]{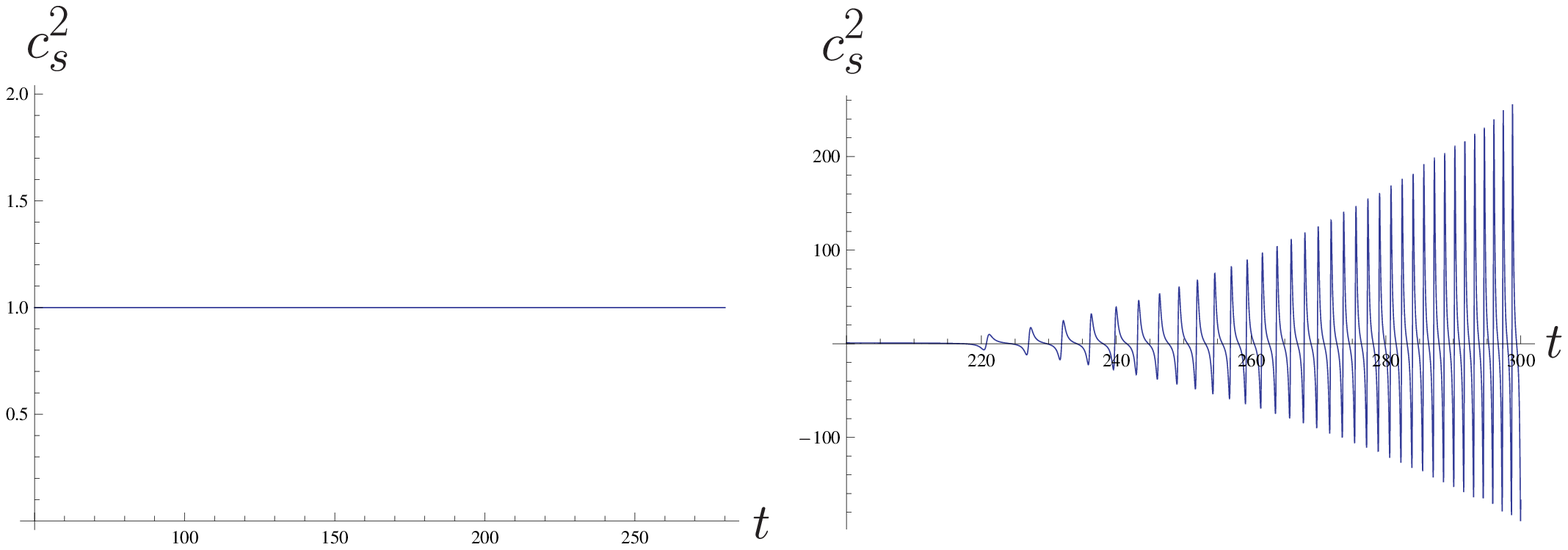}
\end{tabular}
\end{center}
\caption{Sound speed squared $c_s^2$ for curvature perturbation
$\zeta$ after the end of inflation: Left (CC) is constant  and right
(NDC)
 oscillates increasingly  after the end of inflation. }
\end{figure}
%%%%%%%%%%%%%%%%%%%%%%%%%%%%%%%%%%%%%%%%%%%%%%%%%%%%%%%%%%%%%%%%%
%%%%%%%%%%%%%%%%%%%%%%%%%%%%%%%%%%%%%%%%%%%%%%%%%%%%%%%%%%%%%%%%%
%%%%%%%%%%%%%%%%%%%%%%%%%%%%%%%%%%%%%%%%%%%%%%%%%%%%%%%%%%%%%%%%%
%%%%%%%%%%%%%%%%%%%%%%%%%%%%%%%%%%%%%%%%%%%%%%%%%%%%%%%%%%%%%%%%%

On the other hand, it is known that in CC case, the curvature
perturbation $\zeta$ diverges during reheating when
$\dot\phi=0$~\cite{Finelli:1998bu,Jedamzik:2010dq,Easther:2010mr}.
Furthermore, it is apparent  that in NDC case, $\zeta$ is  divergent
when $\dot\phi=\pm\phi_c$ [$\epsilon_N^H=1$] as well as $\dot\phi=0$
[$\epsilon_N^H=0$] during reheating. To see this more closely, we
write  equation of $\zeta$ from the action (\ref{s2f}) as
\begin{eqnarray}\label{ddzeta}
\ddot{\zeta}+[3H+F(\epsilon_N^H)]\dot{\zeta}-\frac{c_s^2}{a^2}\partial^2\zeta=0,
\end{eqnarray}
where
\begin{eqnarray}
F(\epsilon_N^H)=\frac{\dot{\epsilon}_N^H}{\epsilon_N^H}\times\frac{(\epsilon_N^H)^3-3(\epsilon_N^H)^2+7\epsilon_N^H+3}{(\epsilon_N^H+1)(\epsilon_N^H-1)(\epsilon_N^H-3)}.
\end{eqnarray}
We observe that  $F$ behaves as
\begin{eqnarray}\label{exF}
F\simeq\left\{\begin{array}{ll}
\frac{1}{\epsilon_N^H}\simeq\frac{1}{\dot{\phi}^2},
~~({\rm at}~\dot\phi=0)\\
\frac{1}{\epsilon_N^H-1}\simeq\frac{1}{\dot\phi^2-\phi_c^2} ~~({\rm
at}~\dot\phi=\pm\phi_c)\end{array}\right.
\end{eqnarray}
which implies that equation (\ref{ddzeta})  becomes singular either
at $\dot{\phi}=0$ [$\epsilon_N^H=0$] or at $\dot{\phi}=\pm\phi_c$
[$\epsilon_N^H=1$].

However, these singular behaviors must be  checked  at the solution
level in the superhorizon limit. For this purpose, we  consider
Fourier mode $\zeta_k$ and then,  equation (\ref{ddzeta}) becomes
\begin{eqnarray}\label{ddzetak}
\ddot{\zeta}_k+[3H+F(\epsilon_N^H)]\dot{\zeta}_k+\frac{c_s^2k^2}{a^2}\zeta_k=0.
\end{eqnarray}

In the case of $k^2/a^2\gg 1$, equation (\ref{ddzetak}) reduces to
\begin{eqnarray}\label{ddsub}
\ddot{\zeta}_k+\frac{c_s^2k^2}{a^2}\zeta_k\simeq0.
\end{eqnarray}
Since $c^2_s$ oscillates in Fig. 9, the curvature perturbation
$\zeta_k$ leads to an exponential destabilization at small scales,
which is called the gradient instability in the subhorizon.

 For $k^2/a^2\ll 1$, the  superhorizon mode
$\zeta_k$ could be illustrated by
\begin{equation}
\zeta_k(t)\simeq  \label{chi-sol0}
\zeta^{(0)}_k+c_k\int^{\infty}_tdt'\frac{H^2(t')}{A_1^2(t') A_2(t')
a^3(t')},
\end{equation}
where $\zeta^{(0)}_k$ and $c_k$ are constants which are determined
by choosing vacuum and  time of the horizon-crossing $t_k$. Here,
the constant mode $c_k$ is not safe. A correction to the
superhorizon mode  up to $k^2$-order leads to
~\cite{Weinberg:2005vy}
\begin{eqnarray}\label{chi-sol1}
\zeta_k&\simeq&\zeta_{k}^{(0)}\Bigg[1-k^2\int_{t}^\infty
dt'\frac{H^2(t')}{A_1^2(t') A_2(t') a^3(t')}
\int^{t'}_{-\infty}dt''c_s^2(t'')a(t'')\frac{A_1^2(t'')
A_2(t'')}{H^2(t'')}\Bigg].
\end{eqnarray}
Plugging  $A_1 $, $A_2$ in (\ref{a12}) and $c_s^2$ in (\ref{csss})
together with $A_3$ (\ref{a3e}) into the mode (\ref{chi-sol1}), the
first and second integrals of the last term in (\ref{chi-sol1}) are
given by
\begin{eqnarray}
\int_t^\infty dt'\frac{H^2}{A_1^2 A_2 a^3}&=&\int_t^\infty dt'\frac{1}{\epsilon_N^H}\frac{(1-\epsilon_N^H)^2}{(1-\epsilon_N^H/3)(1+\epsilon_N^H)a^3}\nonumber\\
&=&\frac{2M_{\rm P}^2\tilde{M}^2}{3}\int_t^\infty
dt'\frac{1}{\dot{\phi}^2}\frac{[1-3\dot{\phi}^2/(2M_{\rm
P}^2\tilde{M}^2)]^2}{[1-\dot{\phi}^2/(2M_{\rm
P}^2\tilde{M}^2)][1+3\dot{\phi}^2/(2M_{\rm
P}^2\tilde{M}^2)]a^3}\label{int1}
\end{eqnarray}
and
\begin{eqnarray}
\int^{t'}_{-\infty} dt''c_s^2a\frac{A_1^2 A_2}{H^2}&=&\int^{t'}_{-\infty} dt''\Bigg\{\frac{d}{dt''}\Big[\frac{a(1-\epsilon_N^H/3)^2}{H(1-\epsilon_N^H)}\Big]-a\Big(1+\frac{\epsilon_N^H}{3}\Big)\Bigg\}\nonumber\\
&=&\frac{a[1-\dot{\phi}^2/(2M_{\rm
P}^2\tilde{M}^2)]^2}{H[1-3\dot{\phi}^2/(2M_{\rm
P}^2\tilde{M}^2)]}\Bigg|^{t'}_{-\infty}-\int^{t'}_{-\infty}
dt''a\Big(1+\frac{\dot{\phi}^2}{2M_{\rm P}^2\tilde{M}^2}\Big),
\label{int2}
\end{eqnarray}
respectively. Substituting  (\ref{int1}) and (\ref{int2}) into
(\ref{chi-sol1}) leads to
\begin{eqnarray} \label{zetas2}
\zeta_k&\simeq&\zeta_{k{(0)}}\Bigg[1-k^2\frac{2M_{\rm P}^2\tilde{M}^2}{3}\Bigg\{\int_{t}^{\infty}dt'\frac{1}{\dot{\phi}^2}
\frac{[1-\dot{\phi}^2/(2M_{\rm P}^2\tilde{M}^2)][1-3\dot{\phi}^2/(2M_{\rm P}^2\tilde{M}^2)]}{[1+3\dot{\phi}^2/(2M_{\rm P}^2\tilde{M}^2)]a^2H}\nonumber\\
&&\hspace*{3em}-\int_{t}^{\infty}
dt'\frac{1}{\dot{\phi}^2}\frac{[1-3\dot{\phi}^2/(2M_{\rm
P}^2\tilde{M}^2)]^2}{[1-\dot{\phi}^2/(2M_{\rm
P}^2\tilde{M}^2)][1+3\dot{\phi}^2/(2M_{\rm P}^2\tilde{M}^2)]a^3}\times\nonumber\\
&&\Bigg(\frac{a(-\infty)[1-\dot{\phi}^2(-\infty)/(2M_{\rm
P}^2\tilde{M}^2)]^2}{H(-\infty)[1-3\dot{\phi}^2(-\infty)/(2M_{\rm
P}^2\tilde{M}^2)]}+\int_{-\infty}^{t'}
dt''a\Big(1+\frac{\dot{\phi}^2}{2M_{\rm
P}^2\tilde{M}^2}\Big)\Bigg)\Bigg\}\Bigg]
\end{eqnarray}
which implies  that the integrand in (\ref{zetas2})
 diverges at $\dot{\phi}=0$ [$\epsilon_N^H=0$], while it is finite at
$\dot{\phi}=\pm\phi_c$ [$\epsilon_N^H=1$]. We note that even though
the singular behavior at $\dot{\phi}=\pm \phi_c$ disappears at the
solution level, one cannot avoid the blow-up of the curvature
perturbation $\zeta_k$ when $\dot{\phi}=0$. This means that $\zeta$
is unphysical and thus,  one has to reanalyze the perturbation
during the reheating by looking for a physical
gauge~\cite{Germani:2015plv}. This is the Newtonian gauge.

Finally, we would like to mention that the homogenous evolution of
CC+NDC  is not affected by the CC-term in Section 3, provided the
coefficient $\sigma_C$ is taken to be a small value. Since we were
carrying out the perturbation during the NDC-background evolution,
it is not clear how the CC-term influences the perturbation
equations. Hence one should check if the evolution by this term
could be neglected in the perturbation analysis. In order to see it,
one relevant quantity is the sound speed squared $c^2_s$ because it
may show a difference of the evolution between the NDC-dominant case
of CC+NDC and the NDC. As was shown in Eq.(\ref{ddzetak}), this
quantity plays the important role in the perturbed equation for the
curvature perturbation mode $\zeta_k$. From Fig. 9,  we remind the
reader that $c^2_s$ is constant for CC, while it oscillates for NDC.
We have computed $c^2_s$ from the CC+NDC and depicted in Fig. 10. In
this expression, one  term of $\frac{\sigma_C\dot{\phi}^2}{2M^2_{\rm
P}}$ is added to $A_2$ in defining $c^2_s$ (\ref{csss}) while
keeping the remaining unchanged. Comparing Fig. 9 with Fig. 10 shows
that CC-picture [NDC-picture] of $c^2_s$ are very similar to
CC-dominant picture [NDC-dominant picture] $c^2_s$ of CC+NDC. It
indicates that the oscillating behavior  of $c^2_s$ from the NDC
persists in the NDC-dominant $c^2_s$ of CC+NDC. Hence we may neglect
the CC-term in the perturbation analysis of the NDC.
%%%%%%%%%%%%%%%%%%%%%%%%%%%%%%%%%%%%%%%%%%%%%%%%%%%%%%%%%%%%%%%%%
%%%%%%%%%%%%%%%%%%%%%%%%%%%%%%%%%%%%%%%%%%%%%%%%%%%%%%%%%%%%%%%%%
%%%%%%%%%%%%%%%%%%%%%%%%%%%%%%%%%%%%%%%%%%%%%%%%%%%%%%%%%%%%%%%%%
\begin{figure}[t!]
\begin{center}
\begin{tabular}{cc}
\includegraphics[width=.9
\linewidth,origin=tl]{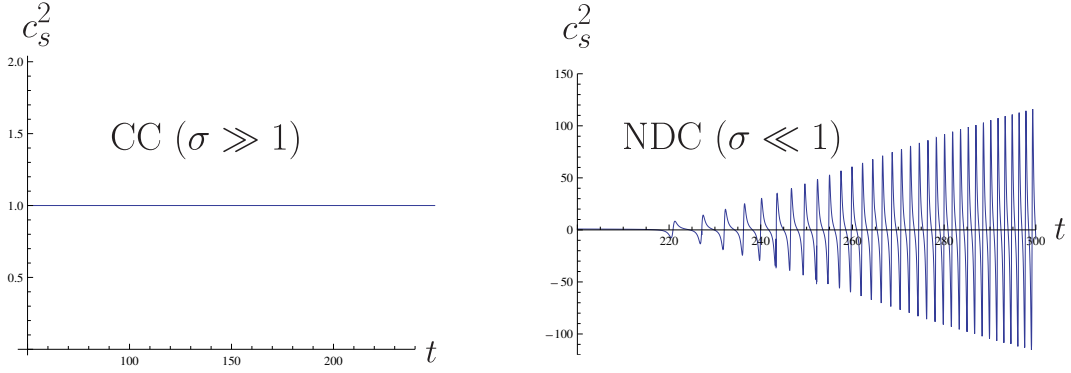}
\end{tabular}
\end{center}
\caption{Sound speed squared $c_s^2$ for curvature perturbation
$\zeta_k$ after the end of inflation: Left (CC-dominant case of
CC+NDC) is nearly constant and right (NDC-dominant case of CC+NDC)
 oscillates increasingly  after the end of inflation. }
\end{figure}
%%%%%%%%%%%%%%%%%%%%%%%%%%%%%%%%%%%%%%%%%%%%%%%%%%%%%%%%%%%%%%%%%
%%%%%%%%%%%%%%%%%%%%%%%%%%%%%%%%%%%%%%%%%%%%%%%%%%%%%%%%%%%%%%%%%
%%%%%%%%%%%%%%%%%%%%%%%%%%%%%%%%%%%%%%%%%%%%%%%%%%%%%%%%%%%%%%%%%
%%%%%%%%%%%%%%%%%%%%%%%%%%%%%%%%%%%%%%%%%%%%%%%%%%%%%%%%%%%%%%%%%

\section{Perturbation analysis in the Newtonian gauge }
As was shown in the previous section, the comoving gauge was not
suitable for analyzing the perturbation during the reheating. This
is so because the curvature perturbation $\zeta$ blows up at
$\dot\phi=0$ on superhorizon scales. We have to re-analyze the
perturbations by choosing a different gauge without problems at
$\dot{\phi}=0$ \cite{Germani:2015plv}.  To this end, we consider the
scalar perturbation around the background
($\phi=\phi(t)+\varphi(t,{\bf x}))$ and the Newtonian
gauge~\cite{Mukh}. Then, the cosmological metric takes the form
\begin{eqnarray}
ds^2_{\rm NG}=-(1+2\Psi)dt^2+a^2(t)(1-2\Phi)dx^i dx^j \delta_{ij}.
\end{eqnarray}
Here $\Psi$ is the Newtonian  potential, while $\Phi$  is the
Bardeen potential~\cite{Motta:2013cwa}. We note  that $\Psi=\Phi$ in
the CC, but for the Horndenski theories including the NDC, $\Psi$ is
not the same with $\Phi$~\cite{Motta:2013cwa}. It is instructive to
note that the Bellini-Sawicki parametrization~\cite{Bellini:2014fua}
is very useful to describe the perturbation compactly on
superhorizon scales including the reheating period. It turns out
that for the NDC model (\ref{mact}), the Newtonian potential $\Psi$
is related to $\Phi$  as
\begin{eqnarray}\label{psieq}
\Psi=\Phi(1+\alpha_{\rm T})\left[1-\frac{\alpha_{\rm M} -\alpha_{\rm
T}}{\epsilon+\alpha_{\rm M}-\alpha_{\rm T}}\right]
-\frac{\alpha_{\rm M}-\alpha_{\rm T}}{H[\epsilon+\alpha_{\rm M}
-\alpha_{\rm T}]}\dot{\Phi}
\end{eqnarray}
with $\epsilon=-\dot{H}/H^2$. Considering the NDC model
(\ref{mact}), one has $K=V=V_0\phi^2,~G_4=M^2_{\rm
P}/2,~G_5=-\phi/2\tilde{M}^2$, which determine the two parameters
$\alpha_{\rm M}$ and $\alpha_{\rm T}$ as
\begin{eqnarray}
\alpha_{\rm M}=-\frac{\dot{\phi}\ddot{\phi}}{H(\tilde{M}^2M_{\rm P}^2
-\dot{\phi}^2/2)},~~~~~\alpha_{\rm T}=
\frac{\dot{\phi}^2}{\tilde{M}^2M_{\rm P}^2-\dot{\phi}^2/2}.
\end{eqnarray}
Also, for the NDC model, the Hamiltonian constraint on superhorizon
scales reduces to
\begin{eqnarray} \label{con-law}
\partial_t \left(\frac{HQ}{\dot{\phi}}\right)=0,
\end{eqnarray}
where $Q$ is the Mukhanov-Sasaki variable (the gauge-invariant
combination)
\begin{equation}
Q=\varphi+\frac{\dot{\phi}}{H}\Phi.
\end{equation}
Eq.(\ref{con-law}) could be recast in terms of the Bardeen potential
$\Phi$
\begin{eqnarray}\label{phieq}
\dot{\Phi}+(1+\epsilon+\alpha_{\rm M})H\Phi=CH[\epsilon+\alpha_{\rm M}
-\alpha_{\rm T}],
\end{eqnarray}
where the constant $C$ depends on the initial conditions $\zeta_c$
settled during inflation when changing the Newtonian gauge to the
comoving gauge.
%%%%%%%%%%%%%%%%%%%%%%%%%%%%%%%%%%%%%%%%%%%%%%%%%%%%%%%%%%%%%%%%%
%%%%%%%%%%%%%%%%%%%%%%%%%%%%%%%%%%%%%%%%%%%%%%%%%%%%%%%%%%%%%%%%%
%%%%%%%%%%%%%%%%%%%%%%%%%%%%%%%%%%%%%%%%%%%%%%%%%%%%%%%%%%%%%%%%%
\begin{figure}[t!]
\begin{center}
\begin{tabular}{cc}
\includegraphics[width=.9
\linewidth,origin=tl]{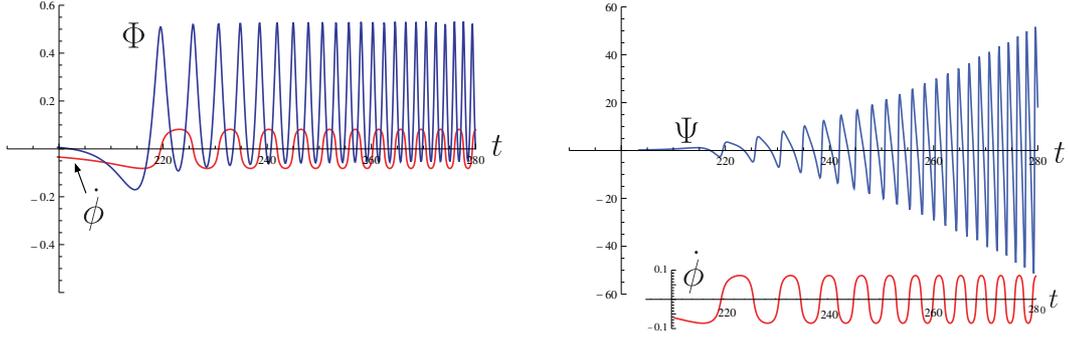}
\end{tabular}
\end{center}
\caption{The behaviors of $(\Phi,\dot{\phi})$ [left] and
$(\Psi,\dot{\phi})$ [right] with respect to time $t$, after the end
of inflation. Both figures show that the evolutions for $\Phi$ and
$\Psi$ are regular at $\dot{\phi}=0$.}
\end{figure}
%%%%%%%%%%%%%%%%%%%%%%%%%%%%%%%%%%%%%%%%%%%%%%%%%%%%%%%%%%%%%%%%%
%%%%%%%%%%%%%%%%%%%%%%%%%%%%%%%%%%%%%%%%%%%%%%%%%%%%%%%%%%%%%%%%%
%%%%%%%%%%%%%%%%%%%%%%%%%%%%%%%%%%%%%%%%%%%%%%%%%%%%%%%%%%%%%%%%%

In the CC+NDC model, one has
$K=V=\lambda\phi^4/4,~G_3=-\phi/2,~G_4=M_{\rm
P}^2/2,~G_5=-\phi/2M^2$~\cite{Germani:2015plv}, where they have
shown that the curvature perturbations $\zeta$  on superhorizon
scales are not generally  conserved but the the rescaled
Mukhanov-Sasaki variable is conserved, implying a constraint
equation for the Newtonian potential. This implies that the
superhorizon perturbations of $\Phi$ and $\Psi$ are fine with the
warning that $\Psi$ could become very large.

Coming back to the NDC model, we solve Eq.(\ref{phieq}) for $\Phi$
numerically by taking into account the reheating period. Also,
making use of Eq.(\ref{psieq}) leads to the numerical solution for
the Newtonian potential $\Psi$.  Explicitly, Fig. 11 shows that
after the end of inflation, the behaviors of $\Phi$ (left) and
$\Psi$ (right) are regular at $\dot\phi=0$. Also, we observe that
the Newtonian potential $\Psi$ grows  very large values.

Since the subhorizon mode $\zeta_k$ of the curvature perturbation
has suffered from the gradient instability in the comoving gauge, it
is important to check whether this instability is present  in the
Newtonian gauge. For this purpose, we use the second-order evolution
equation for the Bardeen potential mode $\Phi_k$ as
\begin{equation}\label{phi-evol}
\ddot{\Phi}_k+\frac{\beta_1\beta_2+\beta_3\alpha^2_{\rm B}
\frac{k^2}{a^2}}{\beta_1+\alpha^2_{\rm B} \frac{k^2}{a^2}}
\dot{\Phi}_k + \frac{\beta_1\beta_4+\beta_1\beta_5
\frac{k^2}{a^2}+c^2_s\alpha^2_{\rm B}
(\frac{k^2}{a^2})^2}{\beta_1+\alpha^2_{\rm B}
\frac{k^2}{a^2}}\Phi_k=0,
\end{equation}
where the parameters $\beta_i(\alpha_i,H)$ were defined in the
Appendix B of Ref. \cite{Bellini:2014fua}. Here we observe the
oscillating sound speed squared $c^2_s$ (\ref{csss}) shown in Fig.
9. This equation was derived by eliminating the inflaton of
$\varphi/\dot{\phi}$ which is not an observable in the Newtonian
gauge. In the subhorizon regime of $k^2/a^2\gg1$,
Eq.(\ref{phi-evol}) reduces to
\begin{equation}\label{phi-evol2}
\ddot{\Phi}_k+(3+\alpha_{\rm M})H\dot{\Phi}_k +
\Big(\frac{\beta_1\beta_5}{\alpha^2_{\rm B}}
+c^2_s\frac{k^2}{a^2}\Big)\Phi_k\simeq0,
\end{equation}
which is rewritten by introducing the Compton mass scale $k_{\rm C}$
 $ [k^2_{\rm C}c^2_s/a^2\equiv\beta_1\beta_5/\alpha^2_{\rm
B}]$~\cite{DeFelice:2010aj} as
\begin{equation}\label{phi-evol3}
\ddot{\Phi}_k+(3+\alpha_{\rm M})H\dot{\Phi}_k +
c^2_s\Big(\frac{k^2_{\rm C}}{a^2}
+\frac{k^2}{a^2}\Big)\Phi_k\simeq0.
\end{equation}
In the case of $k^2/a^2\gg k^2_{\rm C}/a^2,(3+\alpha_{\rm M})H$, the
evolution equation (\ref{phi-evol3}) takes the form
\begin{equation}\label{phi-evol4}
\ddot{\Phi}_k+ c^2_s \frac{k^2}{a^2}\Phi_k\simeq0,
\end{equation}
which leads to the gradient instability for the oscillating $c_s^2$.
However, we would like to mention that in this case, the gradient
instability emerges when taking the extreme quasi-static limit of
the dynamics ($k\to \infty$) in the Newtonian gauge. This contrasts
to the case of the comoving gauge where one could find the gradient
instability easily when requiring the condition of $k^2/a^2\gg1$ as
was shown in Eq.(\ref{ddsub}).

\section{Summary and Discussions}
%%%%%%%%%%%%%%%%%%%%%%%%%%%%%%%%%%%%%%%%%%%%%%%%%%%%%%%%%%%%%%%%%
%%%%%%%%%%%%%%%%%%%%%%%%%%%%%%%%%%%%%%%%%%%%%%%%%%%%%%%%%%%%%%%%%
%%%%%%%%%%%%%%%%%%%%%%%%%%%%%%%%%%%%%%%%%%%%%%%%%%%%%%%%%%%%%%%%%
 First of all, we have studied the difference between NDC and CC during
reheating after the end of inflation.  We have observed a sizable
difference that the inflaton velocity $\dot{\phi}$ oscillates with
damping for CC, while it oscillates without damping for NDC. We have
confirmed that this difference has arisen from  different time rates
of their Hubble parameters ($\dot{H}$).   Analytic expressions for
inflaton and Hubble parameter obtained by applying the averaging
method to the NDC-equations
(\ref{Heq})-(\ref{seq})~\cite{Ghalee:2013ada} are not suitable for
describing violent oscillations of Hubble parameter. Hence their
argument of disappearing the parametric resonance is not proven for
the NDC.

Now, we  mention the perturbative feature for  NDC generated during
reheating. We have studied  the curvature perturbation $\zeta$ by
taking the comoving gauge ($\varphi=0$). This gauge is definitely
applicable at the stage of inflation, but it may be  incompatible
with $\dot{\phi}=0$ during the
 reheating. As was shown in Eq.(\ref{ddsub}) in the subhorizon regime ($k^2/ a^2\gg1$),  the Lagrangian
instability (gradient instability) arises easily because the sound
speed squared $c^2_s$ oscillates during the
 reheating. This presumed instability has arisen  because the authors
in~\cite{Ema:2015oaa} have neglected the second term of
(\ref{ddzetak}). However, this is not true for the case of the
superhorizon limit ($k^2/a^2\ll1$) as was shown in (\ref{chi-sol0}).
 Also, this instability never occurs even for the
correction to the superhorizon mode up to $k^2$-order [see
(\ref{chi-sol1})]. But this case  is meaningless  since  the second
term of (\ref{ddzetak}) is singular at $\dot{\phi}=0,\pm\phi_c$.
Here, it is
 noted that the apparent singular behavior at
$\dot{\phi}=\pm \phi_c$ disappeared at the solution level.

Importantly, it is desirable  to comment on the  incompatibility of
the comoving gauge ($\varphi=0$) with $\dot{\phi}=0$ during the
reheating  in the NDC model. We remind the reader   that the blow-up
of $\zeta$ at $\dot{\phi}=0$ happens because the comoving gauge  is
not suitable for describing the oscillating period, especially for
$\dot{\phi}=0$. This indicates that the curvature perturbation is
not considered as a physical variable, describing  a relevant
perturbation during the reheating.
 Hence, it should not be used to draw any physical conclusion. Here
the Bardeen potential $\Phi$ and Newtonian potential $\Psi$  have
been employed as  physical perturbations by choosing  the Newtonian
gauge. The superhorizon perturbations are fine with the warning that
the Newtonian potential may become large. Finally, we note that the
gradient instability of the Bardeen potential mode $\Phi_k$ appeared
when taking the extremal quasi-static limit of the dynamics ($k \to
\infty$) in the Newtonian gauge and thus, the NDC model would become
unviable in the reheating period.

% \vspace{0.25cm} {\bf Acknowledgement}

%\vspace{0.25cm}
% This work was  supported by the National
%Research Foundation of Korea (NRF) grant funded by the Korea
%government (MEST) (No.2012-R1A1A2A10040499).

\end{document}